\documentclass{article}
\usepackage{ijcai21}

\usepackage{times}

\usepackage{soul}
\usepackage{url}
\usepackage[hidelinks]{hyperref}
\usepackage[utf8]{inputenc}
\usepackage[small]{caption}
\usepackage{graphicx}
\usepackage{amsmath}
\usepackage{booktabs}
\usepackage{caption}
\usepackage{subcaption}
\usepackage{color, colortbl}

\newcommand\kjf[1]{\textcolor{black}{#1}}
\makeatletter
\newcommand*\titleheader[1]{\gdef\@titleheader{#1}}
\AtBeginDocument{%
	\let\st@red@title\@title
	\def\@title{%
		\bgroup\normalfont\large\centering\@titleheader\par\egroup
		\vskip1.5em\st@red@title}
}
\makeatother

\usepackage{latexsym}

\title{Incorporating Deception into CyberBattleSim for Autonomous Defense}

\titleheader{\emph{IJCAI-21 1\textsuperscript{st} International Workshop on Adaptive Cyber Defense}}

\author{
	Erich C. Walter$^1$\and
	Kimberly J. Ferguson-Walter$^2$\And
	Ahmad D. Ridley$^2$
	\\
	\affiliations
	$^1$Naval Information Warfare Center Pacific\\
	$^2$Laboratory for Advanced Cybersecurity Research\\
	\emails
	erich.c.walter@spawar.navy.mil\\
}

\begin{document}
\maketitle

\begin{abstract}
Deceptive elements, including honeypots and decoys, were incorporated into the Microsoft CyberBattleSim experimentation and research platform~\cite{cyberbattlesim}.  The defensive capabilities of the deceptive elements were tested using reinforcement learning based attackers in the provided capture the flag environment.  The attacker's progress was found to be dependent on the number and location of the deceptive elements.  This is a promising step toward reproducibly testing attack and defense algorithms in a simulated enterprise network with deceptive defensive elements. 
\end{abstract}

\section{Introduction}
Cyber defenders are increasingly at an disadvantage against attackers in the unbalanced cyber domain. Skilled defensive operators are limited, their job is difficult and never ending, and it is nearly impossible to prove they have been successful at preventing or ejecting all attackers. Research that advances the ability for cyber defenses to automatically detect, respond, and remediate malicious activity quickly is critical. 

Researchers \kjf{profess} the need for publicly available cybersecurity datasets to further advance machine learning (ML) and artificial intelligence research in the field, but these static datasets are unlikely to help quickly advance autonomous decision making and response.  To train an autonomous agent to learn to take optimal automated actions to impede and mitigate cyber attacks, a dynamic training environment that provides appropriate feedback based on the selected action is required.  Microsoft has recently released such a simulated environment, called CyberBattleSim~\cite{cyberbattlesim}, based on the OpenAI gym interface~\cite{OpenAIGym}. CyberBattleSim ``provides a high-level abstraction of computer networks and cyber security concepts'', like a \emph{physics engine} for the cyber domain, allowing researchers to create scenarios and insert custom autonomous (offensive or defensive) agents for training and testing.

Cyber attackers often rely on human errors which cannot always be anticipated or corrected.  These errors can be introduced through social engineering, spearphishing, an unenforced policy, or a software bug accidentally included by the software developer. Defenders can also take advantage of the fact that there is often a human behind a sophisticated cyber attack. Cyber deception considers the human aspects of an attacker in order to impede cyber attacks and improve security~\cite{heckman_cyber_2015}, \kjf{which can also translate to advantages against automated attackers}. Cyber deception aims to understand and influence an attacker even after they have already infiltrated a network, and ultimately to delay, deter, and disrupt their attack. While some ML methods for detection in cybersecurity are still working on improving true-positive/false-positive rates, cyber deception technologies can often naturally act as a high-confidence early warning mechanism.  Cyber deception has been gaining traction in the cyber defense community, with many commercial vendors profiting from these techniques~\cite{fidelis2021}\cite{trapxsecurity2017}\cite{countercraft2021}\cite{cybertrap2021}\cite{attivo2021}, however they are often omitted from autonomous defense frameworks, taxonomies, and simulations~\cite{openC2}. In this paper we investigate
incorporating cyber deception into CyberBattleSim and provide initial results using it to explore several important research questions.

\section{Related Work}
We are primarily interested in determining the feasibility of incorporating cyber deception components into an autonomous cyber defense environment.  Our goal is to quantify the effects of deception on a cyber attacker trained using reinforcement learning to achieve its goal within the autonomous cyber defense environment.  

Reinforcement learning (RL) is a promising machine learning approach to support the development of autonomous agents that learn (near) optimal sequential decision making in complex, uncertain environments~\cite{sutton_reinforcement_2017}. Between 2015 and 2020 alone, more than 1000 papers about RL have been published in the top  Machine Learning conferences~\cite{10.1145/2740908.2742839}. The increased availability of open-source learning environments, such as OpenAIGym~\cite{OpenAIGym} and AI Safety Gridworlds~\cite{gridworlds} has provided a baseline to improve RL research across different application domains. Advancements in the cyber domain have lagged behind many others, due both to its complexity, and to a lack of available data or learning environments.

Within the cybersecurity domain, there are numerous research papers discussing the application of RL to attacking or defending network environments, such as corporate enterprise networks, Internet-of-Things (IoT) networks, cyber physical systems, and software-defined networks (SDNs)~\cite{nguyen2020deep}. However, very few researchers incorporate some form of deception into their learning environment that train the RL agents. Often when researchers do incorporate some form of deception, they refer to  \emph{deception} or \emph{deception attacks} from an adversarial machine learning (AML) perspective, when an attacker perturbs observations to exploit the RL algorithm itself. Our use of the term cyber deception is different from those commonly used in the AML domain.

There are few, if any, open-source, cybersecurity-based RL experimentation environments that enable researchers to address the real-world challenges encountered when applying RL to the cybersecurity domain and improve the state-of-art in RL cyber applications~\cite{ridley2021}. With the release of CyberBattleSim environment in April 2021, Microsoft, leveraging the Python-based Open AI Gym interface, has created an initial, abstract simulation-based experimentation research platform to train automated agents using RL~\cite{cyberbattlesim}. However, CyberBattleSim does not incorporate any cyber deception in its  network simulation to potentially impede the autonomous attacker's progress in achieving its goal of taking ownership of network resources by exploiting their vulnerabilities.

Although there is a lack of open-source RL training environments that include cyber deception, there is still a growing body of cybersecurity research that investigates deception as a defense against cyber attacks. Some of the research applies game theoretic principles to analyze the interactions of attackers and defenders to deceive the attackers with deceptive information and mitigate damage from the attack~\cite{anwar_gamesec_2020} \cite{bilinski_gamesec_2020} \cite{fugate2019}. Other research also incorporate deception in the form of moving target defenses and adaptive honeypot management to increase attacker cost and to gather threat information. In the latter case, the deception is leveraged to reduce environment uncertainty to improve the RL policies learned by the defender~\cite{huang2020}. Other cyber deception research has focused on human subjects experiments~\cite{moonraker-2020}, sometimes also including also simulations~\cite{8073405}, but these simulations are rarely publicly available to other researchers. However, this research does not incorporate deceptive decoys which are lightweight and low-interaction relative to honeypots.  Decoys are less expensive for the defender to create and manage than honeypots as well as being less intrusive on normal users and system administrators than moving target defenses~\cite{ferguson-walter_hotsos_2019}. Decoys have been shown to impede and delay attacker forward progress regardless of whether the attacker is aware that deception is being used as a defense~\cite{usenix-2021}. We aim to demonstrate that the advantages of cyber deception are also effective in degrading attacker performance and mitigating damage from cyber attacks within an RL simulation learning environment.

\section{Methodology}
Many different cyber deception techniques and technologies have been proposed and studied throughout the years~\cite{fraunholz2018}.  In this paper, we examine three popular types of deception---honeypots, decoys, and honeytokens---to provide breadth to our inquiry. Honeypots tend to be high-fidelity, high-interaction fake systems which are designed to induce attentional tunneling in the attacker. They generally focus on looking especially interesting or vulnerable to lure an attacker into the honeypot and propel further interactions to gain intelligence about attacker tactics, techniques, and procedures (TTPs) during later stages of the Cyber Kill Chain~\cite{hutchins_intelligence-driven_2011}. Decoy Systems tend to be comprised of many low-fidelity, low-interaction fake systems which are designed to mimic the real systems on a network during earlier stages of the Cyber Kill Chain (i.e., reconnaissance). Honeytokens are fictitious tokens (e.g., honeypassword, honeypatch, honeyfile) which can be placed on real or deceptive machines. If stolen they act as an early warning system for defenders, as well as provide additional information that can help with response and mitigation. Decoys and honeytokens both benefit from attacker's confirmation bias---a tendency to search for or interpret information in a way that only affirms one's preconceptions.  Since most things on computer networks are real, and are not traps, the attackers tend to assume the decoys and honeytokens are real, even when presented with contradictory information. 

CyberBattleSim was devised to research autonomous agents using reinforcement learning in computer networks, but did not include any cyber deception concepts.  We incorporated decoys, honeypots, and honeytokens (honeycredentials and honeylinks) into the codebase to investigate several research questions critical to cyber deception.

\subsection{Research Questions}\label{ques}

With this preliminary work we begin to address the following research questions:
\begin{itemize}
	\item Which type of cyber deception is best-suited to different attacker goals?
	\item What effect does the number of deceptive elements have on the attacker goals?
	\item When the number of possible deceptive elements is limited, what difference does the location they are placed make on the attacker goal?
\end{itemize}

For these research questions we looked at the following attacker goals:
\begin{enumerate}
	\item Network Ownership: The attacker has taken over a specific percentage of real nodes on the whole network.
	\item Key Terrain: The attacker has successfully exploited the most valuable asset on the network.
\end{enumerate}

\subsection{Simulation Details}
\begin{figure*}[!h]
	\includegraphics[width=.7\linewidth]{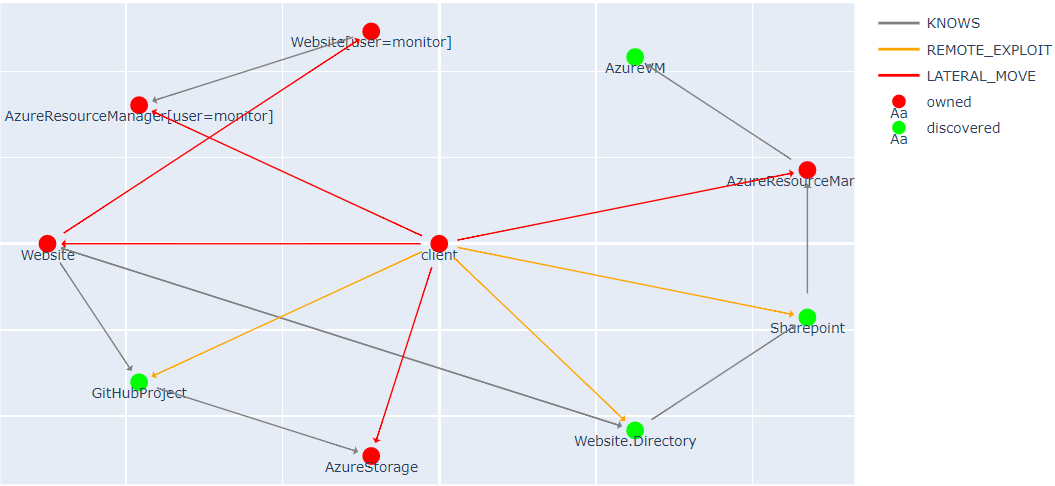}
	\caption{\textbf{CyberBattlerSim Example}: the state space of ToyCTF with no deception.}
	\label{fig:env}
\end{figure*}

The simulation environment was adapted from the toy capture the flag (ToyCTF) example provided in the CyberBattleSim code. The environment is designed to be high-level abstraction of an enterprise computing environment which is represented by a graph where nodes are computing elements and edges are connections between nodes. See Figure ~\ref{fig:env} for an example. We have attempted to maintain variable names from the code in the descriptions used in this paper, for greater transparency, even though they do not always align with the common terminology from the field.

\subsubsection{Action Space }
The action space includes \kjf{local, remote, and connect and control exploits.} \emph{Local exploits} are executable on a node the attacker controls, \emph{remote exploits} are executable on a node the attacker can see, and \emph{connect and control exploits} are executable with a matching credential object on a node visible to the attacker.  Local and remote exploits can reveal undiscovered nodes \kjf{or credentials.  Credentials are used in the \emph{connect and control} action to take over nodes.  }

\subsubsection{State Space and State Transitions }
The scenario examined in ToyCTF involves a small network in which flags are located in a subset of machines, files, or programs (all represented as nodes in the graph). A \emph{local exploit} reveals credentials if there are any on the local node, which can then be used later with a \emph{connect and control exploit} to gain control or ``own'' a specific node, for which that credential is matched. It can also ``discover'' paths to neighboring nodes. A \emph{remote exploit} is similar to the local exploit, but can be used on neighboring nodes.  For the \emph{connect and control exploit} to be successful, 1) the port needs to match, representing the correct service running, 2) the credentials need to match the target node, and 3) the node needs to have been discovered. This action can be taken from any location. The solved capture the flag example provided in the repository was used to infer a successful path for an attacker through the nine nodes in the state space.  

Inferred optimal path from the attacker foothold (Client node): 1) Website, 2) GitHubProject, 3) AzureStorage, 4) Website.Directory, 5) Sharepoint, 6) AzureResourceManager, 7) Website[user=monitor], 8) AzureResourceManager[user=monitor], 9) AzureVM, where [user=monitor] indicates privilege escalation to a user named ``monitor''. This inferred path was used to order the decoy and honeypot additions relative to the start (Client node, where the attacker has gained a foothold on the network) and finish: \emph{early} (nearest the client node), or \emph{late} (nearest the AzureVM node); results are displayed in Figures ~\ref{fig:decoy-location} and ~\ref{fig:honeypots}.  These findings are expected for this scenario and these deceptive elements. The attack path for this scenario entails combining the \emph{Full Ownership}  and \emph{Key Terrain} attacker goals, since the key terrain is at the end of the attack path, and the attacker is required to gain full ownership to progress to the end. The rewards in the ToyCTF example were altered to accommodate the addition of deceptive elements and our new scenario.  

\subsubsection{Deceptive Elements }
The deceptive elements included honeypots~\cite{Mokube_honeypots:concepts}, decoys~\cite{ferguson2017friend}, and honeytokens~\cite{Storey2009HoneyTC} (specifically, honey-credentials and honeylinks).  \emph{Honey-credentials} initially appear to be valid credentials to the attacker but they are invalid and cannot be used to connect to any nodes on the network.
\emph{Honeylinks} reveal potential new nodes, but the nodes are dummy nodes (no penalty for investigating them) and cannot be exploited or taken over.  
Honeypots appear to be a copy of an existing node which can be exploited and controlled.  Honeypots additionally contain a number of honeytokens and honeylinks predefined at the start of the experiment.  Honeypots can be \kjf{exploited by the attacker with the \emph{command and control exploit} but yield no reward}. Decoys appear to be a copy of an existing node, but they can never be exploited or controlled. Credentials for decoys can  be discovered in real nodes, leading the attacker to the decoy, but the connection attempt will fail. Any new action with a honeypot or a decoy generates a defined penalty for the attacker detailed in Section~\ref{sec:reward}.  Repeating actions does not fully penalize the attacker again for the same action, but instead incurs a repeated actions penalty of $-1$.

We have expanded on the ToyCTF example provided within CyberBattleSim which involves a nine node network of real assets. We modeled the following cyber deception techniques within CyberBattleSim:

\begin{itemize}
	\item  \textbf{Decoys}:  cannot be connected to but look like real assets. Connection attempts provide a reward of $-100$, since this will trigger an actionable decoy alert for defenders. No local actions are available on decoys.  Any additional connections to the same decoy garners a $-1$ reward, as a repeated action.
	\item  \textbf{Honeypots}: can be connected to as normal, look like real assets, but are only filled with fake credentials.  The initial successful connection to a honeypot provides a reward of $-100$, since this will trigger a high confidence actionable alert for defenders. Any additional connections to the same honeypot garners a $-1$ reward, as a repeated action.
	\item  \textbf{Honeytokens}: Fake credentials  that do not work on any real machines and  links leading to dummy machines are  located on honeypots. When the attacker uses a honeytoken, it will provide a reward of $-10$, since this will trigger an actionable honeytoken alert for defenders. Repeated use of the same honeytoken garners a $-1$ reward, as a repeated action. Honeylinks also reside within a honeypot, but provide a reward of $0$ to the attacker. They  represent components of honeypots designed to waste attacker time and resources.
\end{itemize}

Honeypots and decoys were generated as clones of existing nodes and connected into the network in parallel to existing nodes with duplicate connections (new actions that target the new node) from the existing node.  These new connections were designed to cancel the reward generated for a successful exploit so there is no reward for exploiting a honeypot.  

\subsubsection{Reward Function }\label{sec:reward}
ToyCTF did not include negative rewards, so the code was changed to allow penalties for interacting with deception.  
\begin{itemize}
	\item Wrong password: -10
	\item Repeated Mistake: -1
	\item Exploit worked: +50
	\item Decoy/honeypot touched: -100
	\item Exploit use: -1
	\item Control of node: +1000
	\item Control of honeypot: 0
	\item Win Condition: +5000
\end{itemize}

We had to modify the  specifics of the default win condition to consider deception. Our win conditions are based on \emph{Network Ownership} and \emph{Key Terrain} described in Section ~\ref{ques}.

\subsubsection{Learning Algorithms }
We utilize the attacker agents provided by CyberBattleSim which include the following algorithms ($\epsilon = 0.9$, epsilon-min = 0.1, epochs = 250, steps = 5000):
\begin{itemize}
	\item \textbf{Credential Lookup (CRED)}: A learner that exploits the credential cache, only taking other actions when no unused credentials are in hand.
	\item \textbf{Deep Q-Learning (DQL)}:  \kjf{A learner that uses a Deep Q-Network (DQN)}. Because it uses Neural Networks as the Q-value function approximator, it has no theoretical convergence guarantees.  It also has Q-value function overestimation bias, like Q-learning, which means the neural network weights/parameters must learn to recover from sub-optimal values.  DQL attempts to overcome stability issues in convergence mainly by using experience replay to store and retrieve previous experiences and delayed updates of neural network weights. 

	\item \textbf{Tabular Q-Learning (QTAB)}, ($\gamma = 0.015$): Theoretically proven to converge to optimal solution under some mild assumptions; however, it is known to overestimate the Q-value function (i.e., state-action value function) because its initial values are randomly chosen at beginning of episodes and it attempts to maximize value at each step.  Convergence of the Q-learning algorithm can be severely slow as it encounters and recovers from local/sub optimal values as a consequence of this overestimation bias.
	\item \textbf{Random Policy (RAND)}: An agent that selects a valid action randomly at each time step; with few available valid actions at each state, this agent performs surprisingly well.
\end{itemize}

For very large state and action spaces, Deep Q-Learning generally outperforms Q-learning, although both overestimate the Q-value function which causes convergence issues.  For smaller states spaces, as in our case, Q-learning tends to converge slowly to a (near) optimal solution, while Deep Q-Learning might not converge at all under certain learning rate conditions. The presence and location of decoys might exacerbate these algorithm traits.  For our purposes, there is no active defender agent. This allows us to better control and compare the attacker learned policies.

\section{Results}
The goal of a RL agent is to learn a policy to maximize expected cumulative reward. For this paper, we used the attacker agents provided in CyberBattleSim.  It may not always be feasible to wrap all critical metrics up into the reward function, so in addition to reward, we consider additional metrics critical for evaluating cyber defense. Additional metrics to consider include \kjf{the number of attacker successes, the amount of resources an attacker wastes due to the deception, and how quickly a defender can detect their presence.}

\begin{figure}
	\centering
	\begin{subfigure}[t]{0.32\textwidth}
		\includegraphics[width=\textwidth]{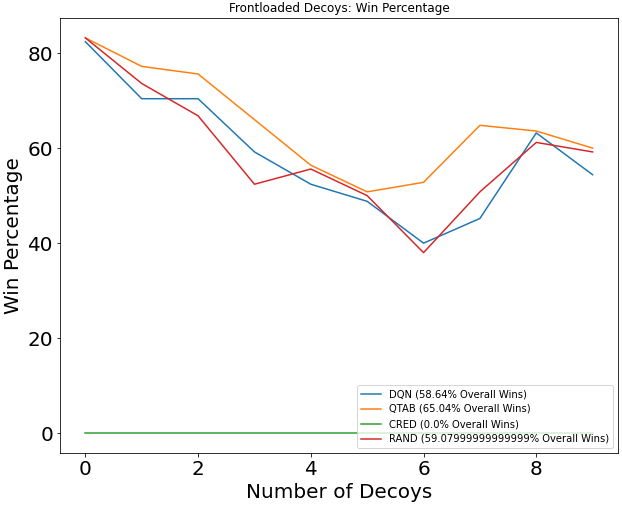}
		\caption{Percentage of attacker wins as the number of decoys increases. DQL: 58.64\%; QTAB: 65.04\%; CRED: 0\%; RAND: 59.08\%}
		\label{fig:newd1}
	\end{subfigure}
	\hfill

	\begin{subfigure}[t]{0.32\textwidth}
		\includegraphics[width=\textwidth]{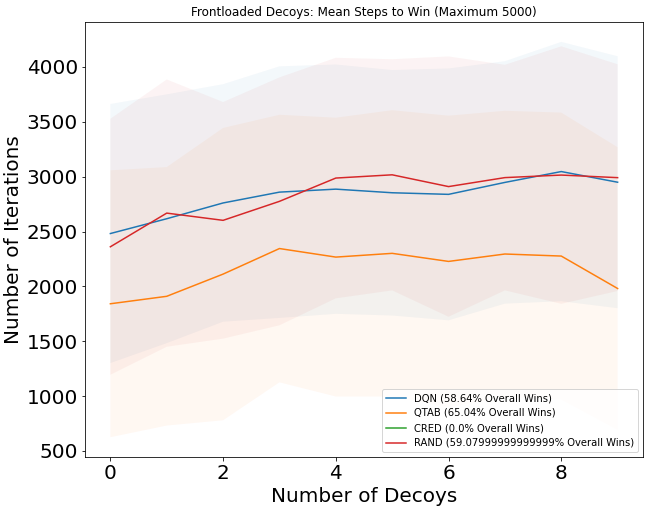}
		\caption{Mean number of steps for attacker to win as a function of the number of decoys. }
		\label{fig:newd1m}
	\end{subfigure}
	\hfill
	\begin{subfigure}[t]{0.32\textwidth}
		\includegraphics[width=\textwidth]{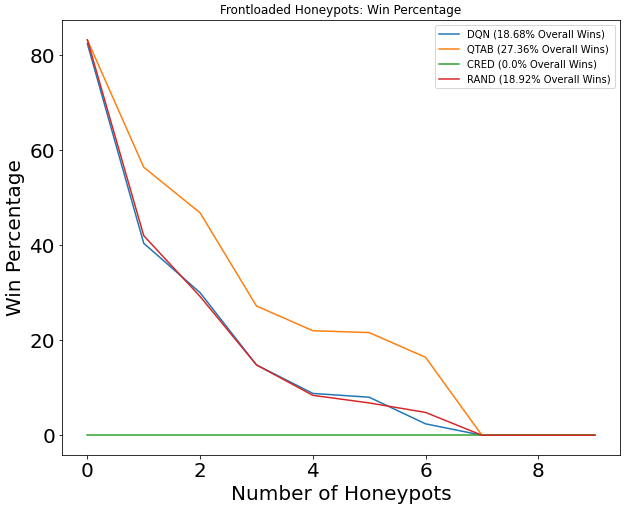}
		\caption{Percentage of attacker wins as number of honeypots increases.  DQL: 18.68\%; QTAB: 27.36\%; CRED: 0\%; RAND: 18.92\%}
		\label{fig:new1}
	\end{subfigure}
	\hfill
	\begin{subfigure}[t]{0.32\textwidth}
		\includegraphics[width=\textwidth]{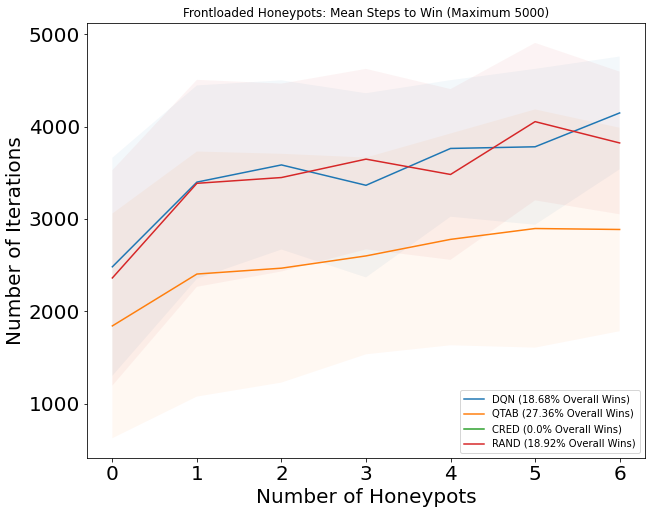}
		\caption{Mean number of steps for the attacker to win as a function of the number of honeypots. }
		\label{fig:new1m}
	\end{subfigure}
	\hfill

	\caption{\kjf{\textbf{Attacker Wins}:  Measures of attacker attaining full ownership when varying the quantity of frontloaded deceptive elements. }}
	\label{fig:attacker-wins}
\end{figure}

\textbf{Attacker Wins} is the percentage of episodes for which the attacker meets with win criteria for the scenario. Figure~\ref{fig:attacker-wins} demonstrates that different attacker agents using different algorithms can have varying overall success on the cyber attack goal. With \kjf{an increasing number of decoys or honeypots,  when \emph{frontloaded} (meaning a decoy of each element in the inferred optimal path was added starting at the beginning)}, the CRED attacker never achieved full ownership, while the agent using QTAB seems to outperform all the others. As expected, honeypots have a greater impact on attacker wins than decoys for each attacker agent, \kjf{and frontloading deceptive elements outperforms \emph{backloading} them; (backloaded graphs not shown due to space limitations).}

\begin{figure*}
	\begin{subfigure}[t]{0.24\textwidth}
	\includegraphics[width=\linewidth]{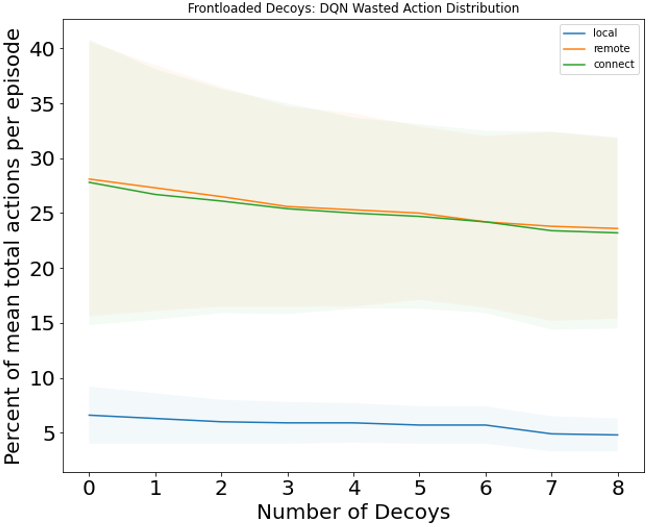}
	\caption{Percentage of mean actions on real nodes as a function of the number of decoys present.}
	\label{fig:w-d}
	\end{subfigure}
	\hfill
		\begin{subfigure}[t]{0.24\textwidth}
		\includegraphics[width=\linewidth]{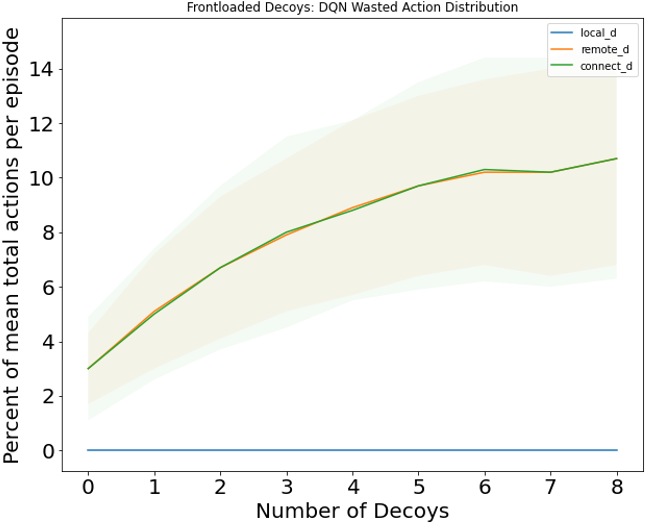}
		\caption{Percentage of mean actions on decoy nodes.}
		\label{fig:s-d}
	\end{subfigure}
	\hfill
	\begin{subfigure}[t]{0.24\textwidth}
	\includegraphics[width=\linewidth]{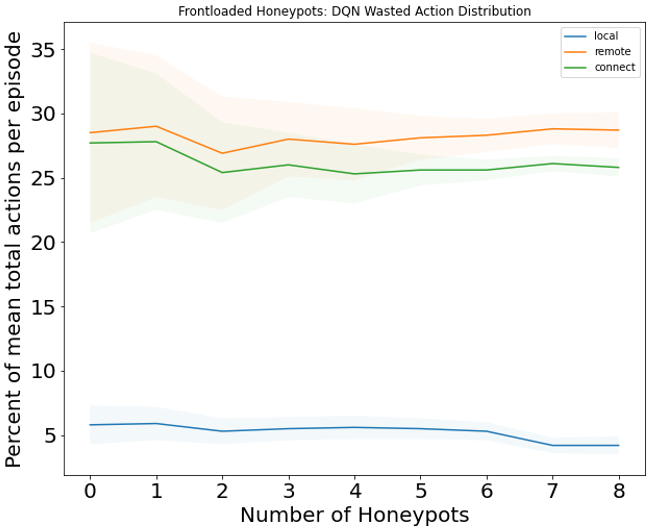}
\caption{Percentage of mean actions on real nodes as a function of the number of honeypots.}
\label{fig:w-h}
	\end{subfigure}
	\hfill
		\begin{subfigure}[t]{0.25\textwidth}
	\includegraphics[width=\linewidth]{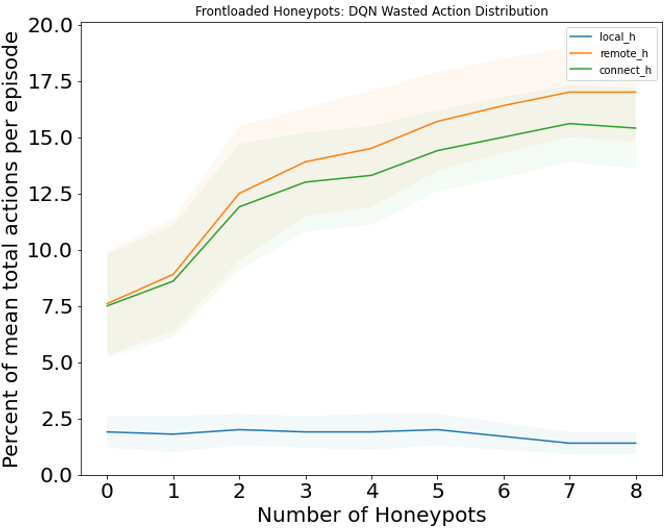}
	\caption{Percentage of mean actions on honeypot nodes.}
	\label{fig:s-h}
\end{subfigure}
\hfill
	\caption{\kjf{\textbf{Wasted Resources Metric}: Percentage of mean actions spent on real or deceptive nodes per episode for the DQL attacker when varying number of a specific type of deceptive element.}}
	\label{fig:attacker-waste}
\end{figure*}

\kjf{\textbf{Wasted Resources} measures the percent of actions the attacker spent on deceptive elements and delaying their true forward progress. As expected, more attacker resources were wasted with the higher fidelity deceptive element---honeypots---and more resources are wasted as the number of deceptive elements increases. Frontloading also had a greater effect. These same patterns are seen across all the experiments (some are not included due to space limitations). These interactions also act as an early warning system, providing high-confidence alerts to the defender from interactions with deceptive elements. This can allow an active defender to pinpoint the attacker’s location, infer their goal, and apply countermeasures. There were also differences in the type of action wasted most among the various attacker agents, though this also depended on features of the deceptive elements (See Figure ~\ref{fig:attacker-waste}). For example, QTAB attacker wasted more \emph{connect and control exploit} actions, less \emph{remote exploit} actions,
and few \emph{local exploit} actions for various quantities of decoys or honeypots, while the number of actions wasted on real nodes (unsuccessful or duplicative actions;
Figure ~\ref{fig:w-d}) decreased slightly as the number wasted on decoys (Figure ~\ref{fig:s-d}) or honeypots increases. This differs from other algorithms, such as DQL, in which \emph{connect and control exploit} actions and \emph{remote exploit} actions seemed to be wasted in equal amounts for decoys. Interestingly, for the case of honeypots, the DQL attacker wasted more \emph{remote exploit} actions, and less \emph{connect and control exploit} actions, and few \emph{local exploit} actions for various quantities of honeypots on both real (Figure ~\ref{fig:w-h}) and deceptive nodes (Figure ~\ref{fig:s-h}). This indicates that just the presence of deceptive elements on the network have potential to alter the attacker’s behavior on the real nodes in the network.}

\kjf{\textbf{Defender Detections} is number of times, and the speed at which, the attacker has triggered a high-confidence alert via deceptive elements, which would allow an active defender to pinpoint the attacker’s location and apply countermeasures. In this case, since both the decoys and honeypots can act as an early warning system, any interaction counts as a detection, so this metric simply depends on the location of the deceptive element in the attack path.}

\subsection{Type of Deception Technology}
These results investigate how each different type of deception affected attacker behavior by comparing the success of the autonomous attackers for each attacker goal. Comparing several of the figures we can compare the deceptive technologies.  The decoys are best at impeding the CRED attacker, since its main TTP (using credentials) is most effected by the design of decoys. This provided agent was designed to ``exploit the credential cache to connect to a node not owned yet''.  The honeypots seem to have a slightly greater effect on the other attacker algorithms, since the multiple honeytokens increase the valid actions available within a honeypot and thus magnify their impact. \kjf{See Figure~\ref{fig:new1} illustrating the reduced attacker success as more honeypots are added to the network (frontloaded), compared to a similar increase in decoys in Figure~\ref{fig:newd1}. However, there are other reasons that a lower fidelity deceptive element (like decoys) may be preferred by defenders, to something more effective (like honeypots).  The reason honeypots are more effective is often because of the enhanced realism and fidelity, allowing longer and more complicated interactions with the attacker. However, this comes as an increased set-up and upkeep cost to defenders to ensure these high-fidelity elements are deep, consistent, and fresh.}\\

\subsection{Quantity of Deceptive Elements}
\kjf{By comparing the success of the included autonomous attackers, we investigate how the quantity of each of the three different types of deception technologies affects attacker behavior.}

\begin{figure*}
	\centering
	\begin{subfigure}[t]{0.33\textwidth}
		\includegraphics[width=\textwidth]{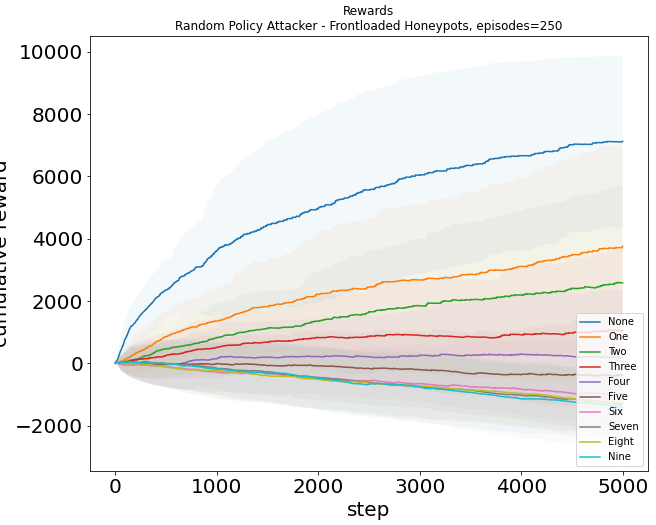}
		\caption{Honeypots Frontloaded RAND Rewards}
		\label{fig:new3}
	\end{subfigure}
	\hfill
	\begin{subfigure}[t]{0.33\textwidth}
		\includegraphics[width=\textwidth]{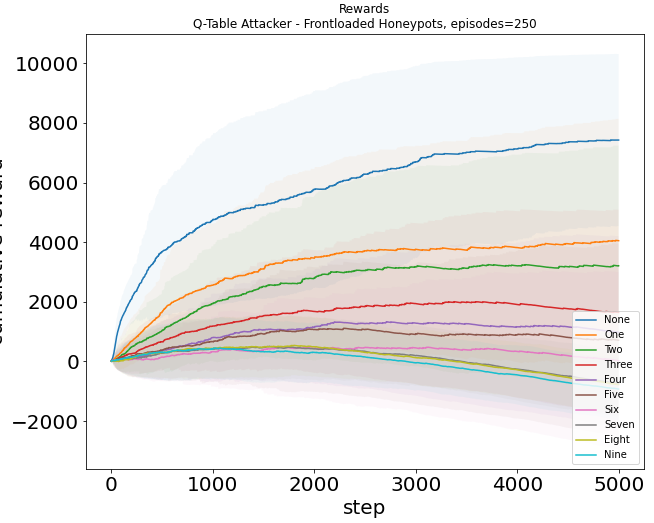}
		\caption{Honeypots Frontloaded QTAB Rewards}
		\label{fig:new5}
	\end{subfigure}
	\hfill
	\begin{subfigure}[t]{0.33\textwidth}
		\includegraphics[width=\textwidth]{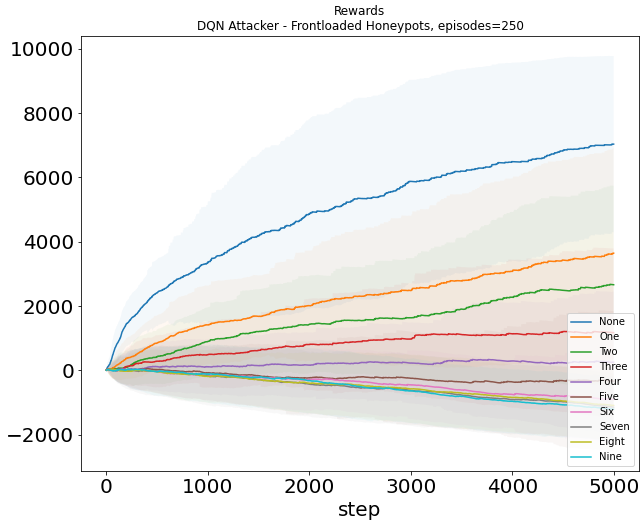}
		\caption{Honeypots Frontloaded DQL Rewards}
		\label{fig:new6}
	\end{subfigure}
	\hfill
	\caption{\kjf{\textbf{Quantity of Honeypots}: Cumulative reward for various attacker agents as a function of the quantity of honeypots (maximum 5000 steps).}}
	\label{fig:honeypots}
\end{figure*}

\begin{figure*}
	\centering
	\begin{subfigure}[t]{0.33\textwidth}
		\includegraphics[width=\textwidth]{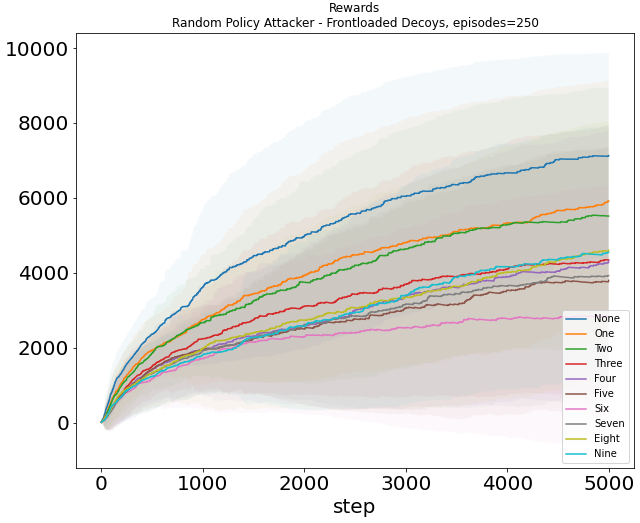}
		\caption{Decoys Frontloaded RAND Rewards}
		\label{fig:new3d}
	\end{subfigure}
	\hfill
	\begin{subfigure}[t]{0.33\textwidth}
		\includegraphics[width=\textwidth]{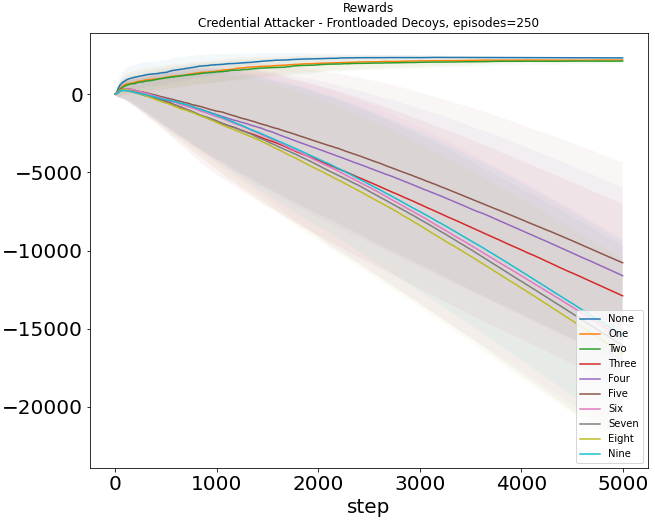}
		\caption{Decoys Frontloaded CRED Rewards}
		\label{fig:new4d}
	\end{subfigure}
	\hfill
	\begin{subfigure}[t]{0.33\textwidth}
		\includegraphics[width=\textwidth]{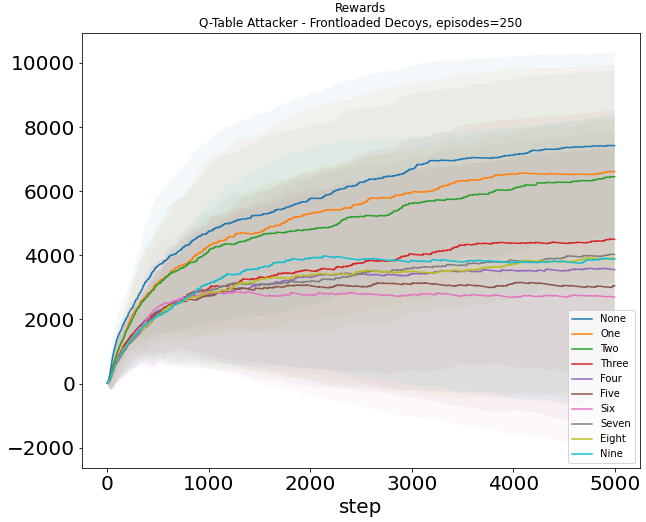}
		\caption{Decoys Frontloaded QTAB Rewards}
		\label{fig:newd5}
	\end{subfigure}
	\hfill
	\begin{subfigure}[t]{0.33\textwidth}
		\includegraphics[width=\textwidth]{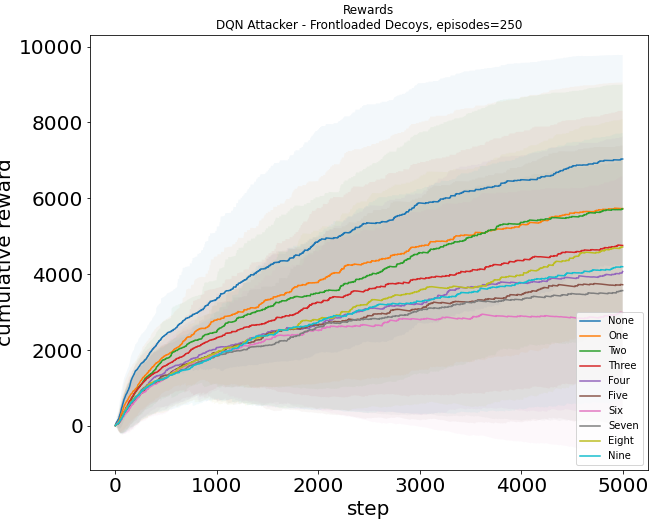}
		\caption{Decoys Frontloaded DQL Rewards}
		\label{fig:new6d}
	\end{subfigure}
					\hfill
		\begin{subfigure}[t]{0.33\textwidth}
			\includegraphics[width=\textwidth]{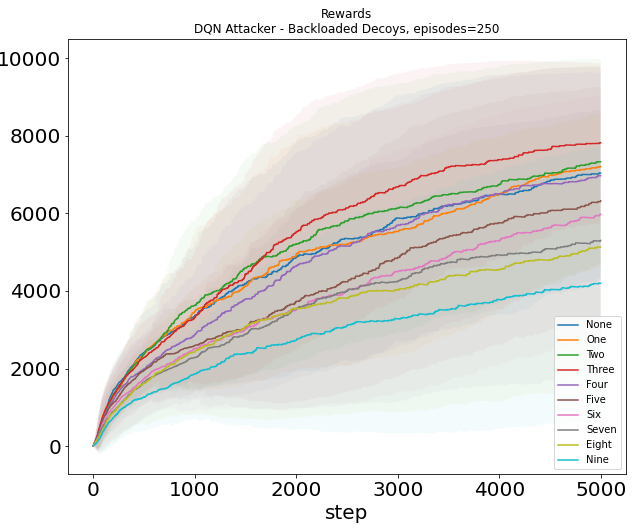}
			\caption{Decoys Backloaded DQL Rewards}
			\label{fig:newd00}
		\end{subfigure}
			\hfill
			\begin{subfigure}[t]{0.33\textwidth}
			\includegraphics[width=\textwidth]{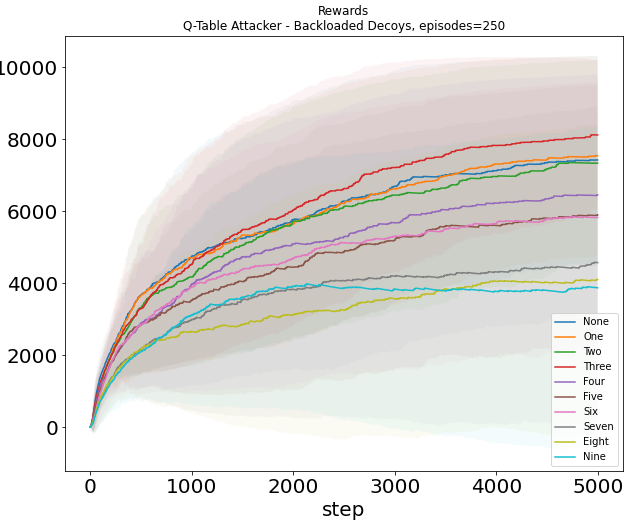}
			\caption{Decoys Backloaded QTAB Rewards}
			\label{fig:newd0}
		\end{subfigure}

	\hfill

	\caption{\kjf{\textbf{Quantity of Decoys}: Cumulative reward for various attacker agents as a function of the quantity of decoys (maximum 5000 steps).}}
	\label{fig:decoys}

\end{figure*}

\textbf{Decoys.} In Figure~\ref{fig:decoys} we examine how the attacker reward changes as the number of decoys is increased.  We compare several default algorithms provided as the number of decoys increases from zero to nine until such there is a decoy copy of every real machine in the network. In the case where there are no decoys, the DQL attacker accumulates the most reward, however, as the number of decoys increases, this attacker's performance dips.  Furthermore, the CRED attacker, who focuses on using credentials to connect to machines, performs much worse as the number of decoys increases. This is because they are trying to connect to the decoy machines with unusable credentials and accumulating negative rewards. \kjf{Examining Figures ~\ref{fig:newd00} and ~\ref{fig:newd0}, compared to Figures ~\ref{fig:new6d} and ~\ref{fig:newd5}, respectively, displays the difference between backloading and frontloading decoys for different automated attackers.}\\
\textbf{Honeypots.} In Figure~\ref{fig:honeypots} we examine how the attacker reward changes as the number of honeypots increases. One honeypot seems to have a similar, but magnified effect of one decoy on the DQL attacker. The more honeypots that are deployed increases the negative effect on the attacker. \kjf{When a honeypot is only encountered at later stages in the attack, this delays the negative effect on attacker reward since no extra barriers were encountered in the earlier part of their campaign.}

\subsection{Layout of Deceptive Elements}
\begin{figure}
	\centering
	\begin{subfigure}[t]{0.35\textwidth}
		\includegraphics[width=\textwidth]{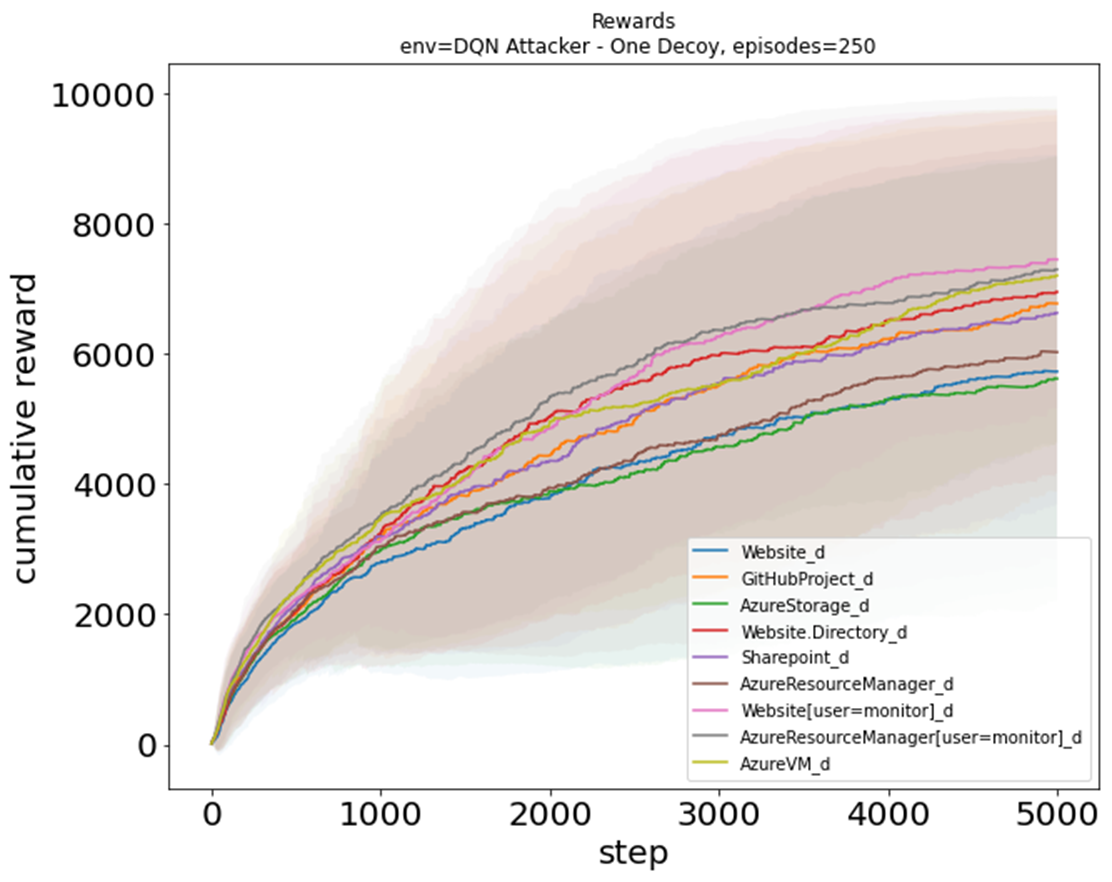}
		\caption{Deep Q-Learning (DQL)}
		\label{fig:DQN-loc-d}
	\end{subfigure}
	\hfill
	\begin{subfigure}[t]{0.35\textwidth}
		\includegraphics[width=\textwidth]{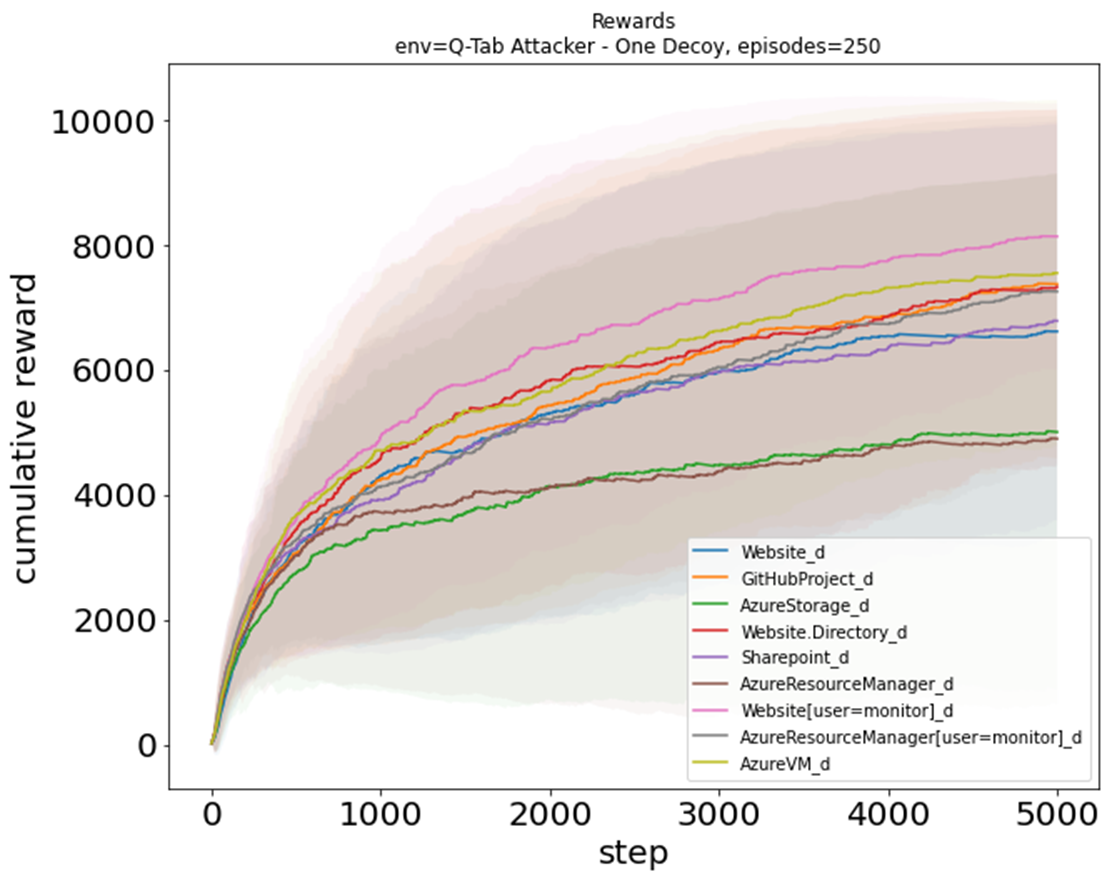}
		\caption{Tabular Q-Learning (QTAB))}
		\label{fig:QTab-loc-d}
	\end{subfigure}
	\hfill
	\begin{subfigure}[t]{0.35\textwidth}
		\includegraphics[width=\textwidth]{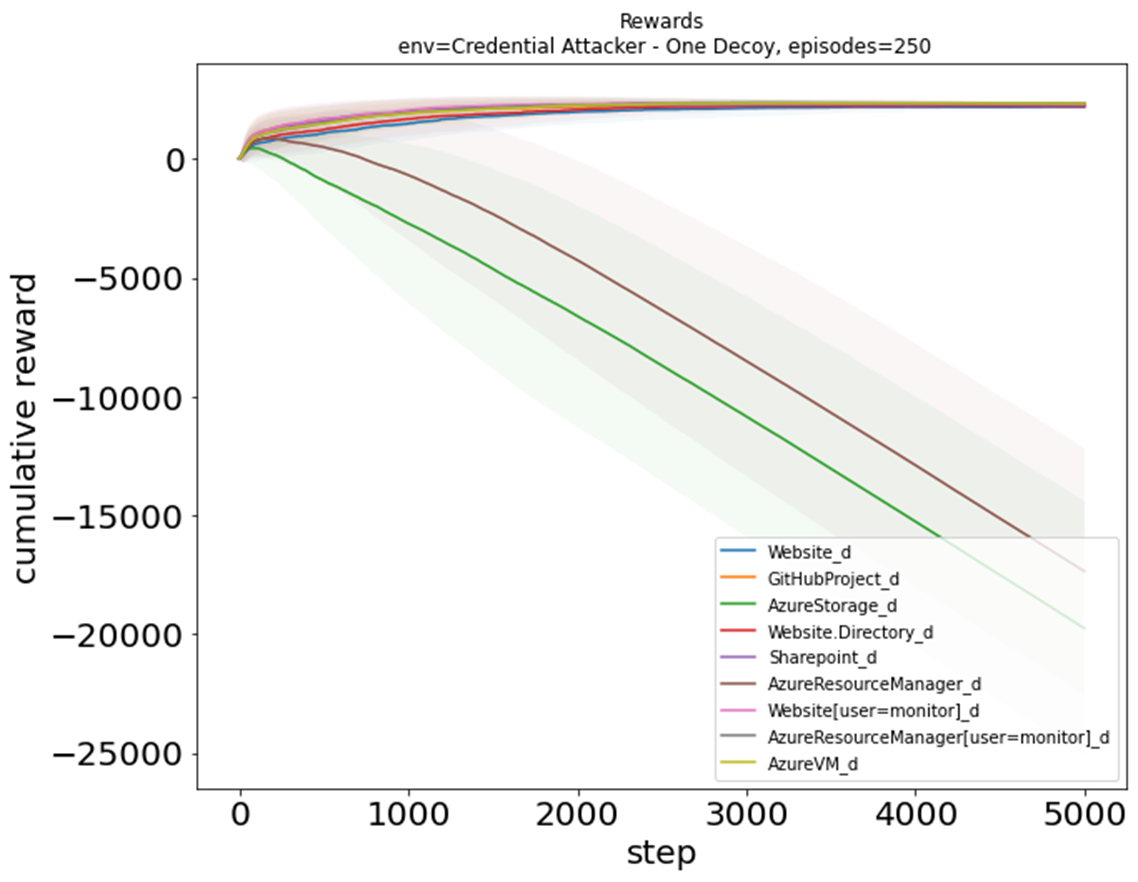}
		\caption{Credential Lookup (CRED)}
		\label{fig:Cred-loc-d}
	\end{subfigure}
	\hfill
	\begin{subfigure}[t]{0.35\textwidth}
		\includegraphics[width=\textwidth]{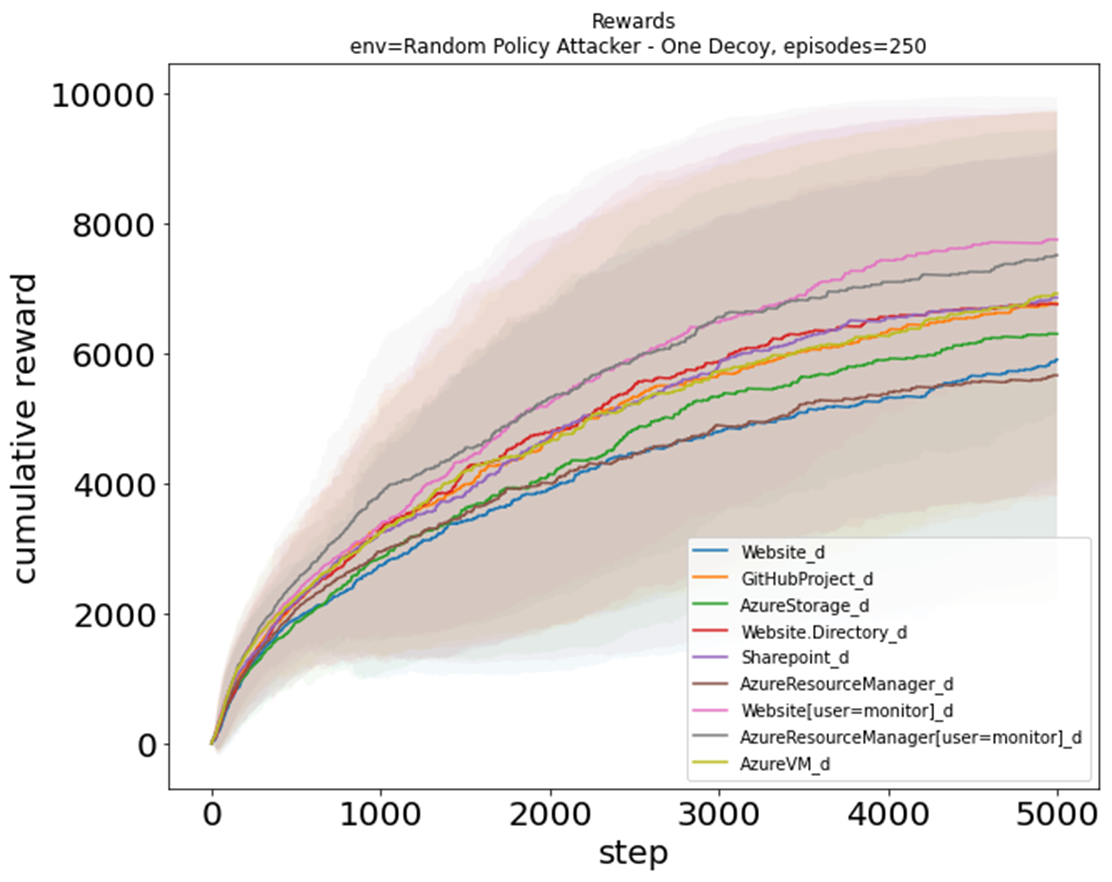}
		\caption{Random Policy (RAND): Agent that selects a valid action randomly at each time step}
		\label{fig:Rand-loc1}
	\end{subfigure}
	\caption{\textbf{Location of Decoys:} Cumulative reward for various attacker agents as a function of location of one decoy. \kjf{Location of the decoy is based on inferred optimal path, with Website at the beginning (step 1) and Azure VM at the end (step 9).}}
	\label{fig:decoy-location}
\end{figure}

\begin{figure}
	\centering
	\begin{subfigure}[t]{0.35\textwidth}
		\includegraphics[width=\textwidth]{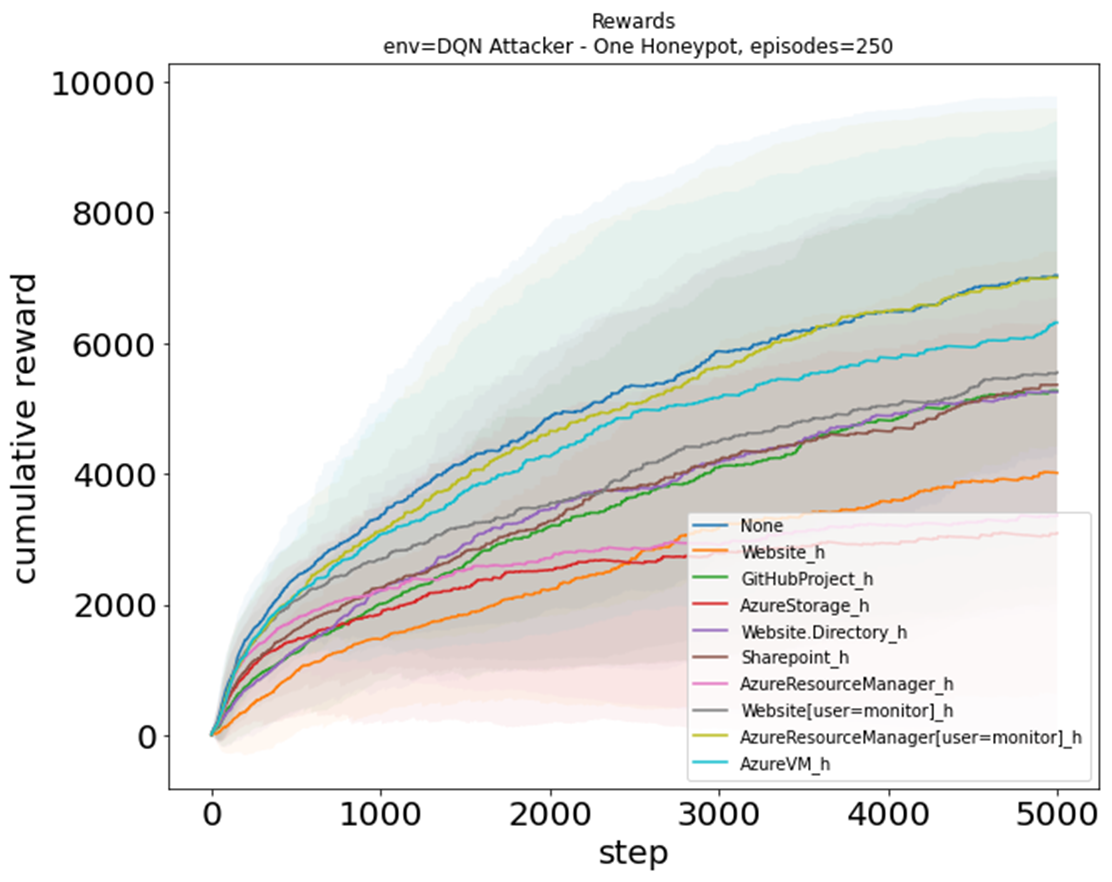}
		\caption{Deep Q-Learning (DQL)}
		\label{fig:DQN-loc-h}
	\end{subfigure}
	\hfill
	\begin{subfigure}[t]{0.35\textwidth}
		\includegraphics[width=\textwidth]{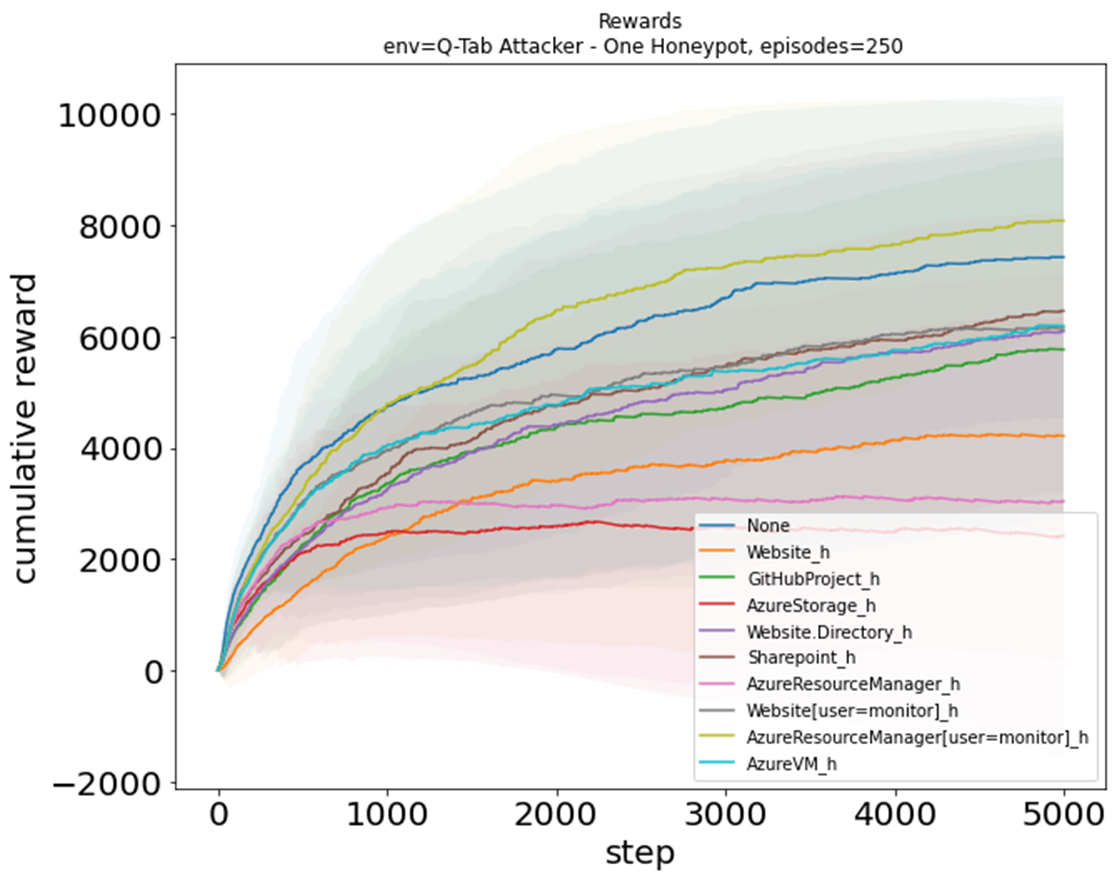}
		\caption{Tabular Q-Learning (QTAB)}
		\label{fig:QTab-loc-h}
	\end{subfigure}
	\hfill
	\begin{subfigure}[t]{0.35\textwidth}
		\includegraphics[width=\textwidth]{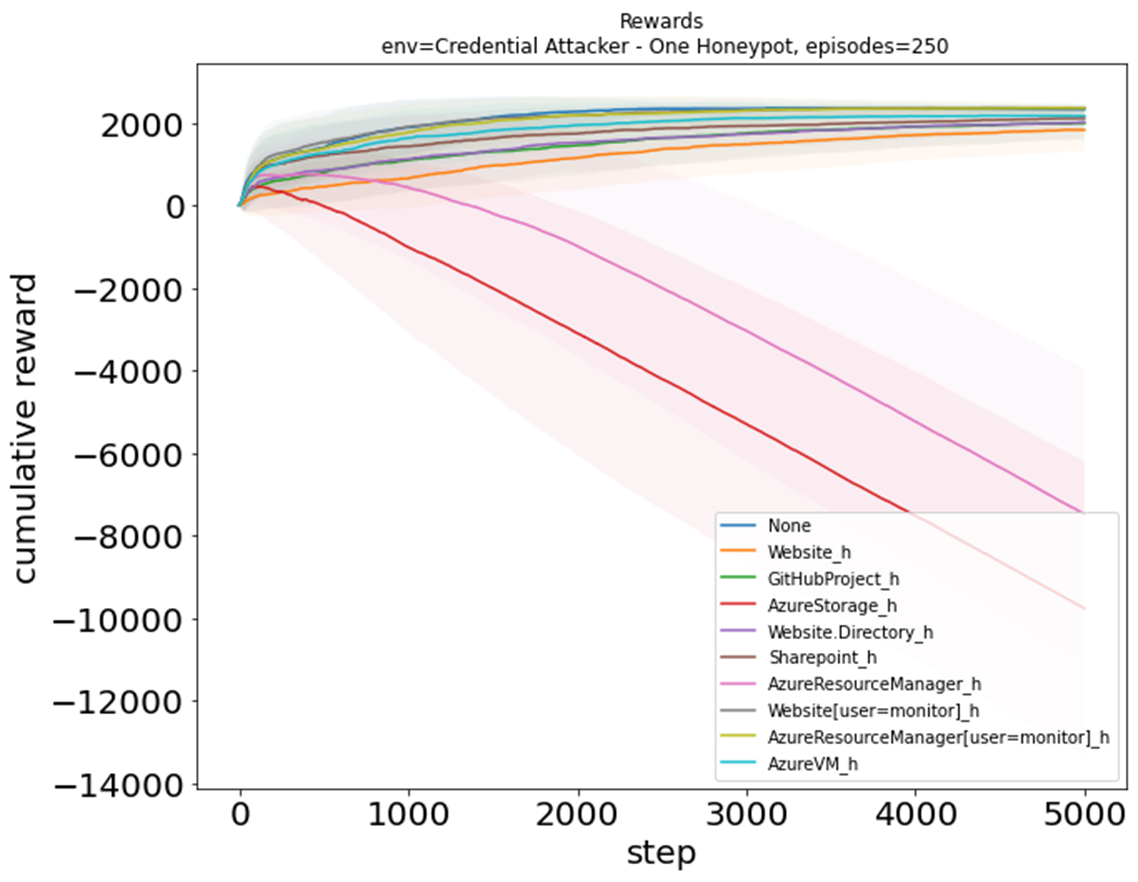}
		\caption{Credential Lookup (CRED)}
		\label{fig:Cred-loc-h}
	\end{subfigure}
	\hfill
	\begin{subfigure}[t]{0.35\textwidth}
		\includegraphics[width=\textwidth]{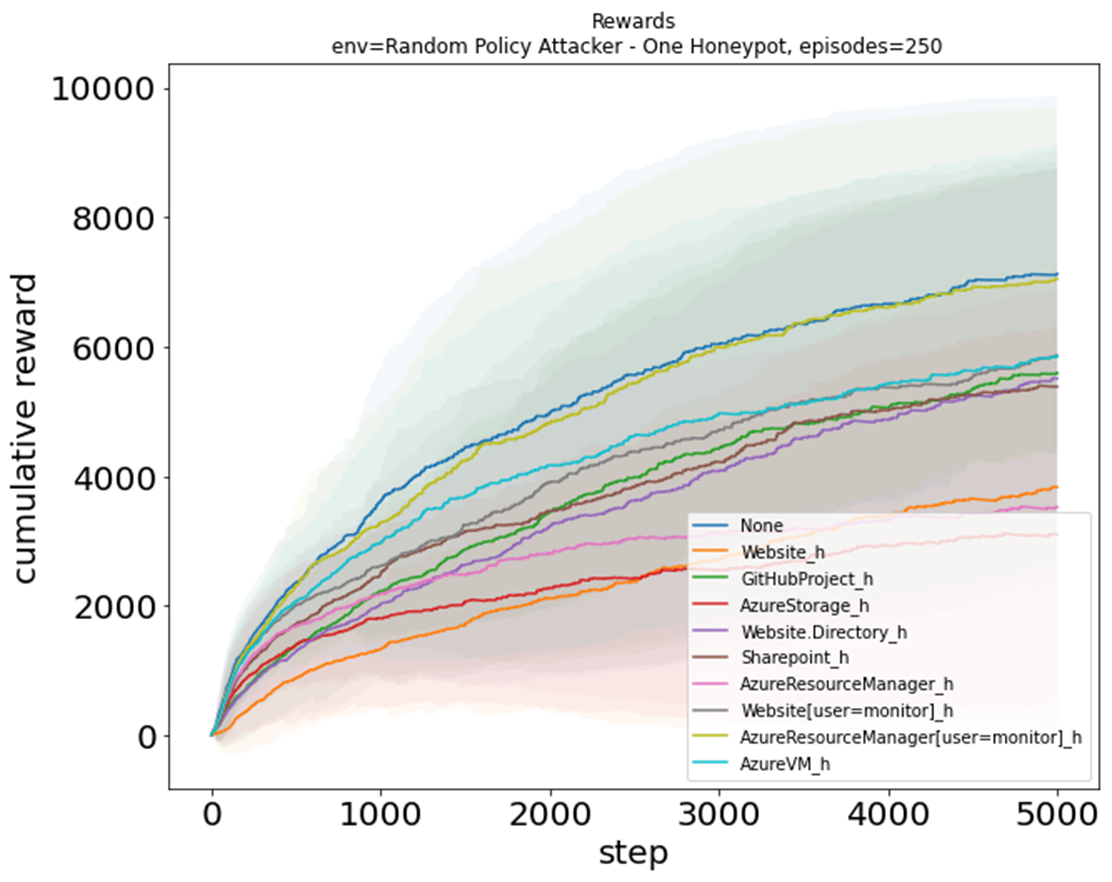}
		\caption{Random Policy (RAND): Agent that selects a valid action randomly at each time step}
		\label{fig:Rand-loc}
	\end{subfigure}
	\caption{\textbf{Location of Honeypots}: Cumulative reward for various attacker agents as a function of the location of one honeypot. \kjf{Location of the decoy is based on inferred optimal path, with Website at the beginning (step 1) and Azure VM at the end (step 9).}}
	\label{fig:honeypot-location}
\end{figure}

These results illustrate how the deployment layout of two different deception technologies (decoys and honeypots) affects attacker behavior \kjf{by comparing the success of the CyberBattleSim autonomous attackers}.  In real-world defensive situations it may not be evident where to deploy deceptive elements (i.e., at a static \emph{frontloaded} or \emph{backloaded} network location) since attackers do not advertise their foothold location or goal.  Using an autonomous defender, we no longer need to rely on static elements.  The defender agent, no longer needing to rely on static elements, can dynamically change the configuration or layout of deceptive elements based on real-time observations and alerts that can provide updated information about the attacker’s locations and intentions.\\
In Figures~\ref{fig:decoy-location} and \ref{fig:honeypot-location} we examine how the attacker reward changes as the location of a single deceptive element changes, assuming the network owner is resource constrained. The \emph{AzureStorage} decoy (location step 3) and the \emph{Azure Resource Manager} decoy (location step 6) impacted the QTAB and CRED, while the DQL agent was also affected by the Website decoy (location step 1). This kind of simulation can help resource-limited network owners or autonomous defenders determine the best placement for deceptive elements.

\subsection{Implications for an Autonomous Defender}
The results presented in this paper, while not employing a defender agent, provide insight for an autonomous defender. A defender can take a response action immediately when a decoy, honeypot, or honeytoken alert is triggered, including ejecting the attacker from the network. \kjf{Early placement of deceptive elements can go beyond a defender using this high-confidence early warning detection system to block or eject an attacker.  It can more directly restrict an attacker's ability to progress and achieve their goals.}  Furthermore, there are many additional actions an autonomous defender can take including deploying additional deceptive elements~\cite{AlShaer2019AutonomousCD}.  Type, quantity, and layout of the deceptive elements are all relevant features that the defender agent can learn to optimize. Our next steps include the addition of a deceptive defensive agent.

\section{Conclusions \& Future Work}
Results presented in this paper demonstrate expected effects of cyber deception on attacker behavior.  \kjf{The colored shading in the figures shows the standard deviation from each line, which in some cases indicates that neighboring lines have overlapping possible results.} While there are no unexpected findings, the contribution to demonstrate the capability of modeling cyber deception in CyberBattleSim was achieved.  These fundamental results provided a necessary sanity check while incorporating deception into CyberBattleSim.  With a reasonable model of cyber deception and how it effects cyber attacks built into the simulator, researchers can take further steps towards designing custom autonomous defender agents that can actively take deceptive responses and running more complex deception experiments with different scenarios and techniques. Additionally, more sophisticated automated attacker agents could be integrated within the environment to test cyber deception~\cite{10.1145/2991079.2991111}.

Future work also entails informing the models and rewards with insights from real-world attack behavior gathered from human subject experiments~\cite{major2021}. This grounding in reality will help ensure that results are generalizable and applicable to real systems and the humans who attack them. 

\bibliographystyle{plain}
\small
\bibliography{bibliography4-old}

\begin{thebibliography}{10}

\bibitem{8073405}
Palvi Aggarwal, Cleotilde Gonzalez, and Varun Dutt.
\newblock Modeling the effects of amount and timing of deception in simulated
  network scenarios.
\newblock In {\em 2017 International Conference On Cyber Situational Awareness,
  Data Analytics And Assessment (Cyber SA)}, pages 1--7, 2017.

\bibitem{AlShaer2019AutonomousCD}
E.~Al-Shaer, Jinpeng Wei, Kevin~W. Hamlen, and Cliff~X. Wang.
\newblock Autonomous cyber deception: Reasoning, adaptive planning, and
  evaluation of honeythings.
\newblock {\em Autonomous Cyber Deception}, 2019.

\bibitem{anwar_gamesec_2020}
Amhed~H. Anwar and Charles Kamhoua.
\newblock Game theory on attack graph for cyber deception.
\newblock In Q.~Zhu, J.S. Baras, R.~Poovendran, and J.~Chen, editors, {\em
  Decision and Game Theory for Security. {(GameSec)}}, volume 12513 of {\em
  Lecture {{Notes}} in {{Computer Science}}}. Springer, 2020.

\bibitem{10.1145/2991079.2991111}
Andy Applebaum, Doug Miller, Blake Strom, Chris Korban, and Ross Wolf.
\newblock Intelligent, automated red team emulation.
\newblock In {\em Proceedings of the 32nd Annual Conference on Computer
  Security Applications}, ACSAC '16, page 363?373, New York, NY, USA, 2016.
  ACM.

\bibitem{attivo2021}
{Attivo Networks, Inc.}
\newblock Deception and {{Denial}} for {{Detecting Adversaries}}.
\newblock
  https://attivonetworks.com/solutions/threat-detection/cyber-deception/.
\newblock Accessed: 2021-03-04.

\bibitem{bilinski_gamesec_2020}
M.~Bilinski, J.~DiVita, K.J. Ferguson-Walter, S.J. Fugate, R.~Gabrys,
  J.~Mauger, and B.J. Souza.
\newblock Lie another day: Demonstrating bias in a multi-round cyber deception
  game of questionable veracity.
\newblock In Q.~Zhu, J.S. Baras, R.~Poovendran, and J.~Chen, editors, {\em
  Decision and Game Theory for Security. {(GameSec)}}, volume 12513 of {\em
  Lecture {{Notes}} in {{Computer Science}}}. Springer, 2020.

\bibitem{OpenAIGym}
Greg Brockman, Vicki Cheung, Ludwig Pettersson, Jonas Schneider, John Schulman,
  Jie Tang, and Wojciech Zaremba.
\newblock {OpenAI Gym}.
\newblock {\em CoRR}, abs/1606.01540, 2016.

\bibitem{countercraft2021}
CounterCraft.
\newblock {{CounterCraft Cloud}}\texttrademark.
\newblock https://www.countercraftsec.com/.
\newblock Accessed: 2021-03-04.

\bibitem{fidelis2021}
Fidelis Cybersecurity.
\newblock Fidelis {Deception}.
\newblock https://fidelissecurity.com/products/deception/.
\newblock Accessed: 2021-03-04.

\bibitem{cybertrap2021}
CYBERTRAP.
\newblock {{CYBERTRAP}}: {{Deception Technology}} from {{Austria}}.
\newblock https://cybertrap.com/en/.
\newblock Accessed: 2021-03-04.

\bibitem{openC2}
{Edited by Jason Romano and Duncan Sparrell}.
\newblock Open command and control {(OpenC2)} language specification version
  1.0.
\newblock
  https://docs.oasis-open.org/openc2/oc2ls/v1.0/cs02/oc2ls-v1.0-cs02.html, Nov
  2019.
\newblock OASIS Committee Specification 02.

\bibitem{ferguson-walter_hotsos_2019}
K.J. Ferguson-Walter, S.J. Fugate, J.~Mauger, and M.M. Major.
\newblock Game theory for adaptive defensive cyber deception.
\newblock {\em ACM Hot Topics in the Science of Security Symposium (HotSoS)},
  March 2019.

\bibitem{ferguson2017friend}
K.J. Ferguson-Walter, D.S. LaFon, and T.B. Shade.
\newblock Friend or faux: deception for cyber defense.
\newblock {\em Journal of Information Warfare}, 16(2):28--42, 2017.

\bibitem{usenix-2021}
K.J. Ferguson-Walter, M.M. Major, C.K. Johnson, and D.H. Muhleman.
\newblock Examining the efficacy of decoy-based and psychological cyber
  deception.
\newblock In {\em USENIX Security Symposuim}, April 2021.

\bibitem{fraunholz2018}
D.~Fraunholz, S.D. Ant{\'o}n, C.~Lipps, D.~Reti, D.Krohmer, F.~Pohl, M.~Tammen,
  and H.~Schotten.
\newblock Demystifying deception technology: A survey.
\newblock {\em ArXiv}, abs/1804.06196, 2018.

\bibitem{fugate2019}
Sunny Fugate and Kimberly {Ferguson-Walter}.
\newblock Artificial {{Intelligence}} and {{Game Theory Models}} for
  {{Defending Critical Networks}} with {{Cyber Deception}}.
\newblock {\em AI Magazine}, 40(1):49--62, March 2019.

\bibitem{heckman_cyber_2015}
K.E. Heckman, F.J. Stech, R.K. Thomas, Be. Schmoker, and A.W. Tsow.
\newblock {\em Cyber {Denial}, {Deception} and {Counter} {Deception}: {A}
  {Framework} for {Supporting} {Active} {Cyber} {Defense}}.
\newblock Advances in {Information} {Security}. Springer International
  Publishing, 2015.

\bibitem{huang2020}
Q.~Huang~L., Zhu.
\newblock Strategic learning for active, adaptive, and autonomous cyber
  defense.
\newblock In Jajodia S., Cybenko G., Subrahmanian V., Swarup V., and Wang~C.
  andWellman M., editors, {\em Adaptive Autonomous Secure Cyber Systems}, Cham,
  2020. Springer.

\bibitem{hutchins_intelligence-driven_2011}
E.M. Hutchins, M.J. Cloppert, and R.M. Amin.
\newblock Intelligence-driven computer network defense informed by analysis of
  adversary campaigns and intrusion kill chains.
\newblock {\em Leading Issues in Information Warfare \& Security Research},
  1(1):80, 2011.

\bibitem{gridworlds}
Jan Leike, Miljan Martic, Victoria Krakovna, Pedro~A. Ortega, Tom Everitt,
  Andrew Lefrancq, Laurent Orseau, and Shane Legg.
\newblock {AI} safety gridworlds.
\newblock {\em CoRR}, abs/1711.09883, 2017.

\bibitem{major2021}
M.M. Major, B.J. Souza, J.~DiVita, and K.J. {Ferguson-Walter}.
\newblock Informing autonomous deception systems with cyber expert performance
  data.
\newblock In {\em International {{Joint Conference}} on {{Artificial
  Intelligence}}: {{Adaptive Cyber Defense Workshop}}}, August 2021.

\bibitem{Mokube_honeypots:concepts}
Iyatiti Mokube and Michele Adams.
\newblock Honeypots: Concepts, approaches, and challenges.
\newblock In {\em in ACM-SE 45: Proceedings of the 45th Annual Southeast
  Regional Conference,}, pages 321--326, 2007.

\bibitem{ridley2021}
A.~Molina-Markham, R.~Winder, and A.~Ridley.
\newblock Network defense is not a game.
\newblock arXiv, 2021.

\bibitem{nguyen2020deep}
Thanh~Thi Nguyen and Vijay~Janapa Reddi.
\newblock Deep reinforcement learning for cyber security.
\newblock arXiv, jul 2020.

\bibitem{trapxsecurity2017}
TrapX Security.
\newblock Active {{Defense}}: {{TrapX}}\textregistered{}
  {{DeceptionGrid}}\textregistered{}.
\newblock
  https://www.forescout.com/company/resources/active-defense-trapx-deceptiongrid-and-forescout/,
  2017.
\newblock Accessed: 2021-03-04.

\bibitem{moonraker-2020}
T.B. Shade, A.V. Rogers, K.J. Ferguson-Walter, S.B. Elsen, D.~Fayette, and K.E.
  Heckman.
\newblock The {Moonraker Study}: {An Experimental Evaluation of Host-Based
  Deception}.
\newblock In {\em Hawaii International Conference on System Sciences (HICSS)},
  Maui, Hawaii, January 2020.

\bibitem{10.1145/2740908.2742839}
Arnab Sinha, Zhihong Shen, Yang Song, Hao Ma, Darrin Eide, Bo-June~(Paul) Hsu,
  and Kuansan Wang.
\newblock An overview of microsoft academic service ({MAS}) and applications.
\newblock In {\em Proceedings of the 24th International Conference on World
  Wide Web}, WWW '15 Companion, pages 243--246. ACM, 2015.

\bibitem{Storey2009HoneyTC}
D.~Storey.
\newblock Honey tokens: Catching flies with honey tokens.
\newblock {\em Network Security archive}, 2009:15--18, 2009.

\bibitem{sutton_reinforcement_2017}
Richard~S. Sutton and Andrew~G. Barto.
\newblock {\em Reinforcement {Learning}: {An} {Introduction}}.
\newblock MIT Press, Cambridge, MA, 2nd edition, 2017.

\bibitem{cyberbattlesim}
Microsoft Defender~Research Team.
\newblock {CyberBattleSim}.
\newblock https://github.com/microsoft/cyberbattlesim, 2021.
\newblock {Created by Christian Seifert, Michael Betser, William Blum, James
  Bono, Kate Farris, Emily Goren, Justin Grana, Kristian Holsheimer, Brandon
  Marken, Joshua Neil, Nicole Nichols, Jugal Parikh, Haoran Wei.}

\end{thebibliography}
\end{document}